# Subwavelength metal grating metamaterial for polarization selective optical antireflection coating

Wonkyu Kim,[1] Junpeng Guo,[1,*] and Joshua Hendrickson[2]

[1] Department of Electrical and Computer Engineering, University of Alabama in Huntsville, Huntsville, AL 35899, USA
[2] Air Force Research Laboratory, Sensors Directorate, WPAFB, Ohio 45433, USA
*Corresponding author: guoj@uah.edu

A metamaterial structure consisting of a one-dimensional metal/air-gap subwavelength grating is investigated for optical antireflection coating on germanium substrate in the infrared regime. For incident light polarized perpendicularly to the grating lines, the metamaterial exhibits effective dielectric property and Fabry-Perot like plasmon-coupled optical resonance results in complete elimination of reflection and enhancement of transmission. It is found that the subwavelength grating metamaterial antireflection structure does not require a deep subwavelength grating period, which is advantageous for device fabrication. Maximal transmittance of 93.4% with complete elimination of reflection is seen in the mid-wave infrared range.

## 1. INTRODUCTION

Metamaterials are artificial materials consisting of subwavelength metal-dielectric structures. Because the feature size of a metamaterial is smaller than the wavelength of interest, a metamaterial can be treated as a spatially-averaged homogeneous material with effective electromagnetic properties determined by the structure and compositions of the unit cell [1, 2]. Metamaterials can be engineered to exhibit electromagnetic properties which cannot be found in nature. Various metamaterial properties, such as negative refractive index [3], high refractive index [4], and optical magnetism [5] have been reported for controlling electromagnetic waves. One kind of metamaterials which exhibit highly anisotropic electromagnetic properties, have been extensively investigated recently [6-14]. This type of metamaterials is called "hyperbolic materials." A hyperbolic metamaterial has dielectric-like properties in one direction and metallic properties in orthogonal directions, or vice versa.

In this paper, we investigate a hyperbolic metamaterial structure for optical antireflection coating on germanium substrate surface in the mid-wave infrared regime. The hyperbolic metamaterial consists of a one-dimensional (1D) metal/air-gap subwavelength grating. The metal/air-gap subwavelength grating exhibits dielectric material properties for polarization perpendicular to the grating lines. By changing the fill factor and the period of the grating, the effective dielectric constant can be designed. Optical wave interference in the dielectric-like metamaterial can reduce optical reflection and enhance optical transmission, similar to traditional quarter-wave dielectric layer anti-reflection coatings. In a traditional quarter-wave antireflection coating, index of refraction of the dielectric layer is required to match the square root of the product of the refractive indices of the substrate material and the incident region. The quarter-wave dielectric anti-reflection coating forms a single mode Fabry-Perot optical cavity. At the resonant wavelength, reflection can be completely suppressed and transmission can be enhanced. For IR anti-reflection applications, choice of dielectric materials is limited because not all dielectric materials are transparent in the infrared regime. In addition, dielectric materials with matched refractive indices for quarter-wave anti-reflection coatings do not always exist in nature. Recently, a strategy of using metamaterial surfaces for anti-reflection has been reported to overcome the index matching requirement in the terahertz and infra-red ranges [15-17]. The structured metal thin films provide designer optical reflection coefficients which can cause destructive interference in the reflection region. In this letter, we investigate an alternative approach for achieving anti-reflection by engineering the effective dielectric constant of a hyperbolic metamaterial to reduce the reflection. In contrast to the 2D antireflection coating structure, the antireflection coating structure in this paper is a 1D structure which consists of only one metamaterial layer between air and the substrate. Furthermore, feature size of the 1D structure is not in the deep-subwavelength regime, which is advantageous for fabrication.

## 2. SUBWAVELENGTH METAL GRATING METAMATERIAL FOR ANTIREFLECTION

The anti-reflective metamaterial is made of a 1-D subwavelength gold metal grating on a germanium substrate. The schematic of the metamaterial structure is shown in Fig. 1. The metal grating has a subwavelength period of $\Lambda$, a metal grating line-width of $w$, and a height of $h$. The fill factor of the metal grating lines is defined as $f = w/\Lambda$. Therefore, the width of metal grating lines can be written as $a = \Lambda f$ and the width of the air-gaps can be written as $b = \Lambda(1 - f)$.

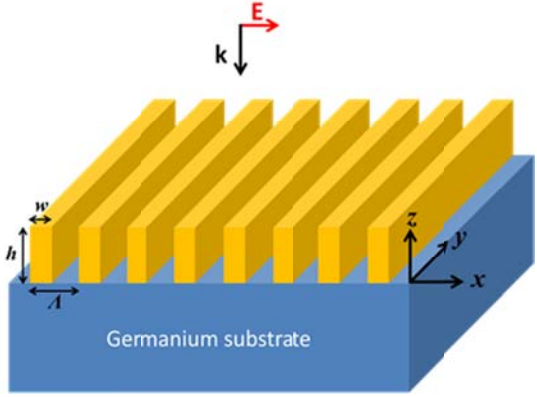

Fig. 1. Schematic of 1D metal subwavelength grating metamaterial structure on a germanium substrate. Incident light is polarized in the $x$ direction and propagates in the $-z$ direction. The period of the grating is $\Lambda$, the metal line-width is $w$, and the height is $h$.

When the period of the subwavelength grating is much smaller than the wavelength, the effective electric permittivities in the x, y, and z directions can be approximately obtained from the following equations [18],

$$\frac{1}{\varepsilon_{xx}} = \frac{f}{\varepsilon_m} + \frac{1-f}{\varepsilon_{air}}, \quad (1)$$

$$\varepsilon_{yy} = \varepsilon_{zz} = f\varepsilon_m + (1-f)\varepsilon_{air}. \quad (2)$$

We calculated the effective electric permittivities versus wavelength for a fill factor of 0.53 and plotted the results in Fig 2. In Fig. 2(a), it can be seen that $\varepsilon_{xx}$ falls in the range from 2.143-0.0020j to 2.128-0.0003j in the wavelength range from 2 to 10 microns. Real parts of $\varepsilon_{yy}$ and $\varepsilon_{zz}$ have negative values, as seen in Fig. 2(b), thus the 1D grating behaviors as a hyperbolic metamaterial with dielectric-like property in the x polarization and metallic properties in the y and z polarizations; i.e. $Re[\varepsilon_{xx}] > 0$ and $Re[\varepsilon_{yy}] = Re[\varepsilon_{zz}] < 0$, over a wide spectral range.

Equations (1) and (2) are valid in the deep subwavelength period regime. For periods not in the deep subwavelength regime, we can calculate the effective permittivity by solving the Bloch wave equation. The dispersion relation between Bloch wave number $K$, period $\Lambda$, and wave numbers $k_{1x}$ and $k_{2x}$ is given in the following equation [18]

$$\cos(K\Lambda) = \cos(k_{1x}a)\cos(k_{2x}b) - \frac{1}{2}\left(\frac{\varepsilon_2 k_{1x}}{\varepsilon_1 k_{2x}} + \frac{\varepsilon_1 k_{2x}}{\varepsilon_2 k_{1x}}\right)\sin(k_{1x}a)\sin(k_{2x}b). \quad (3)$$

In above equation, the Bloch wave number K is the Bloch wave propagation constant in the x-direction. $\varepsilon_1$ and $\varepsilon_2$ are the electric permittivities of gold and air, respectively. For the wave propagating in the -z direction, the Bloch wave vector $K$ is zero. In equation (3), $k_{1x}$ and $k_{2x}$ can be written in terms of $\beta$ in the following equations, where $\beta$ is the Bloch wave propagation constant in the –z direction

$$k_{1x} = \sqrt{\varepsilon_1\left(\frac{\omega}{c}\right)^2 - \beta^2}, \quad k_{2x} = \sqrt{\varepsilon_2\left(\frac{\omega}{c}\right)^2 - \beta^2}. \quad (4)$$

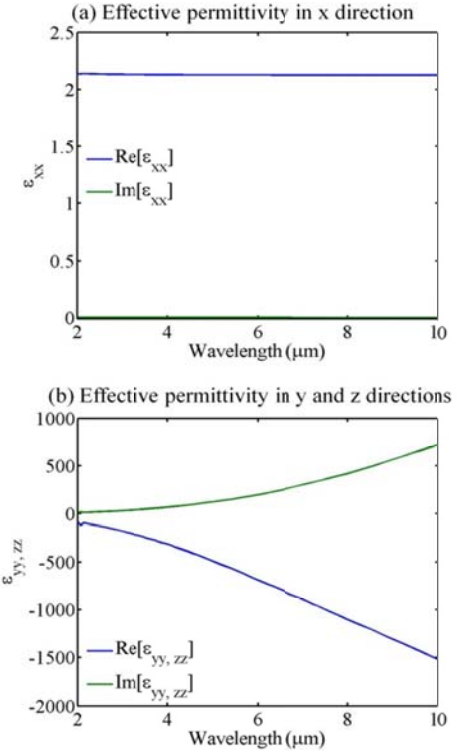

Fig. 2. Effective electric permittivities of a deep subwavelength metal grating in the x direction (a) and in the y and z directions (b). The fill factor of the grating is 0.53.

We numerically solved equation (3) for $\beta$. Then we plotted $(\beta/k_0)^2$ versus the wavelength for different subwavelength grating periods. $k_0$ is the wave number in free space. $(\beta/k_0)^2$ is the effective electric permittivity of the metamaterial. The effective permittivities are shown in Fig. 3. Effective permittivity in the x direction in Fig. 2(a) is shown again for comparison, labeled as the approximation in the Fig. 3. In Fig. 3(a), the real part of the effective permittivity approaches the value of equation (1) as the period approaches to the deep subwavelength regime. When the period increases in the subwavelength regime, the real part of the effective permittivity decreases, approaching to one. As shown in Fig. 3(b), as the period decreases and approaches to the deep subwavelength regime, the imaginary part of the electric permittivity also decreases and approaches to the value of deep subwavelength regime given by equation (1), except for the period of 400 nm. We also observed that the imaginary part of the electric permittivity decreases as the period increases over 200 nm.

Optical transmittance and reflectance from the antireflective metamaterial structure were calculated for various structure parameters using finite-difference time domain (FDTD) simulations with a commercial software code (Lumerical Solutions, Inc.). First, transmittance and reflectance versus wavelength were calculated for different metal fill factors with a fixed grating period of 400 nm and a fixed grating height of 1100 nm. Incident light is polarized in the x direction and propagates in the -z direction. The optical constants of gold and germanium were taken from reference [19]. Fig. 4(a) shows a 2D plot of the transmittance versus wavelength (horizontal axis) and fill factor (vertical axis). Transmittance values higher than 90% are encircled by a white dashed line. The maximal transmittance is 93.2% for a fill factor of 0.53 at 4.66 μm wavelength. Fig. 4(b) shows a 2D plot of reflectance versus wavelength (horizontal axis) and fill factor (vertical axis).

also be considered for easier fabrication. This property is very desirable for fabricating antireflective metamaterial coatings.

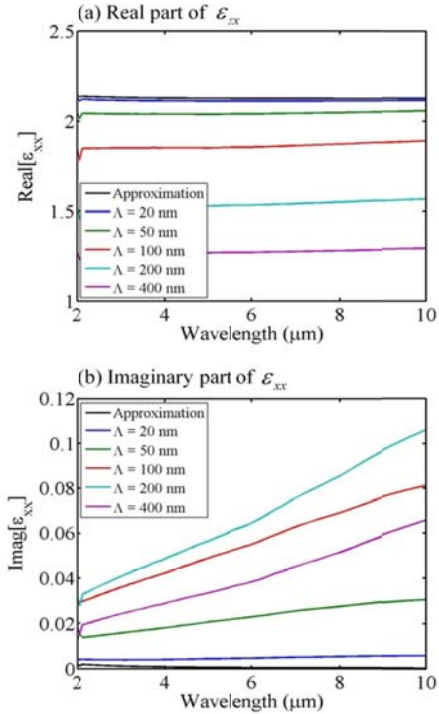

Fig. 3. (a) Real and (b) imaginary part of effective permittivity of the metamaterials for periods of 20, 50, 100, 200, and 400 nm and a fill factor 0.53, along with the deep subwavelength effective permittivity.

Reflectance values lower than 3% are also encircled by a white dashed line. The minimal reflectance is essentially zero (0.005%) for a fill factor of 0.57 at 4.53 µm wavelength. When the fill factor is 0, there is no metal and transmission and reflection occur at the boundary of germanium and air. As the fill factor increases, the real part of the electric permittivity in the x direction remains positive and increases from zero while the imaginary part of the electric permittivity is small. Increasing the fill factor enhances optical wave interference in the structure causing destructive interference in the reflection region. For fill factors in the range from 0.45 to 0.72, reflectance is lower than 3.0%. Correspondingly, when the fill factor falls between 0.42 and 0.64, transmittance is higher than 90%. As the fill factor approaches 1, the metamaterial turns into a uniform gold film which has zero transmission and high reflection. Thus, the optimal fill factor range for high transmission (>90%) and low reflection (<3%) is between 0.45 and 0.64.

Next, we calculated the transmittance and reflectance for different grating periods by fixing the fill factor at 0.53 and the grating height at 1100 nm. Calculation results are shown in Fig. 5. The maximal transmittance of 93.4% occurs for a grating period of 460 nm at the wavelength of 4.57 µm. A minimal reflectance of 0.3% occurs for the grating period of 390 nm at the wavelength of 4.66 µm. Transmittance higher than 90% and reflectance lower than 3% are encircled with white dashed lines in Figs. 5(a) and 5(b), respectively. It can be seen that there is a large range of grating periods, from 180 nm and 680 nm, for which anti-reflection can occur. This implies that hyperbolic metamaterial anti-reflection does not require a deep subwavelength grating period. Although the maximal transmittance occurs at 460 nm, longer period up to 680 nm can

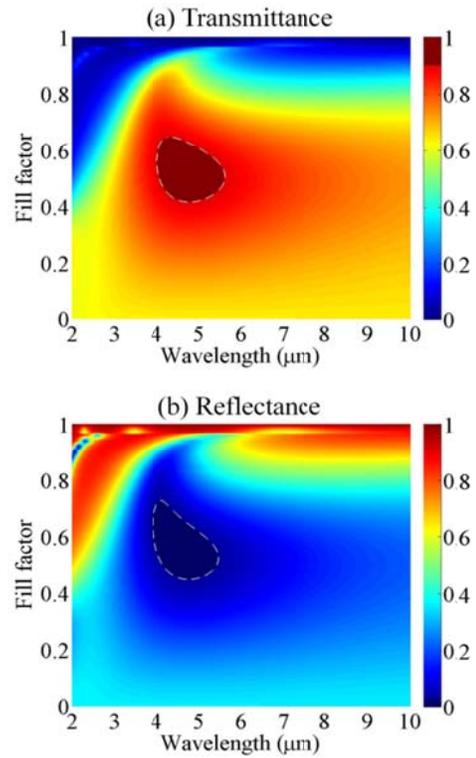

Fig. 4. (a) Transmittance and (b) reflectance versus fill factor and wavelength. Period is 400 nm and grating height is 1.1 µm.

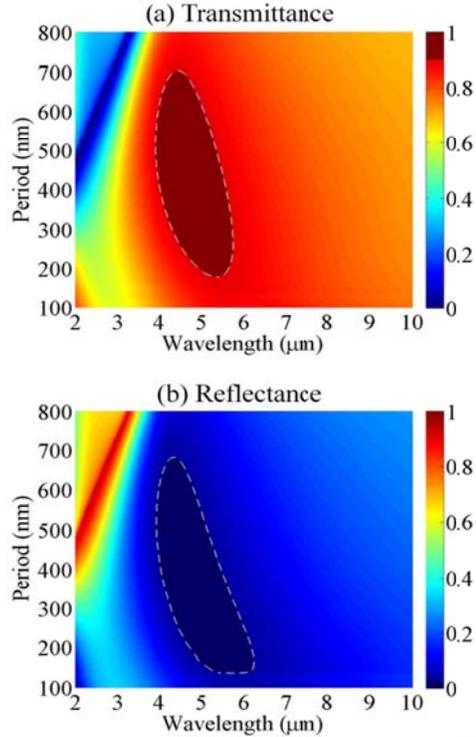

Fig. 5. (a) Transmittance and (b) reflectance versus wavelength for different grating periods. The fill factor is 0.53. The height is 1.1 µm.

Finally, we calculated transmittance and reflectance as a function of grating height by fixing fill factor at 0.53 and grating period at 400 nm. The results are shown in Fig. 6. It can be seen that the peak transmission wavelength scales linearly with the grating height. The linear dependence of the peak transmission wavelength on the height of the grating suggests the occurrence of Fabry-Perot optical cavity interference. Lower and higher bands correspond to the first and second Fabry-Perot resonance modes. Because the first mode exhibits higher transmission and wider bandwidth, the first mode is preferable for antireflection. The peak transmission wavelength is approximately four times of the grating height; i.e., the antireflective metamaterial layer thickness is about one quarter of the free space wavelength. The maximal transmittance was found to be 93.2% at 1110 nm grating height at 4.71 μm wavelength. The minimum of the reflectance is almost zero (0.06%) at a grating height of 2.13 μm at 9.37 μm wavelength. The second order optical resonance enhanced transmission can also be seen in Fig. 6, which occurs for thicker grating layers.

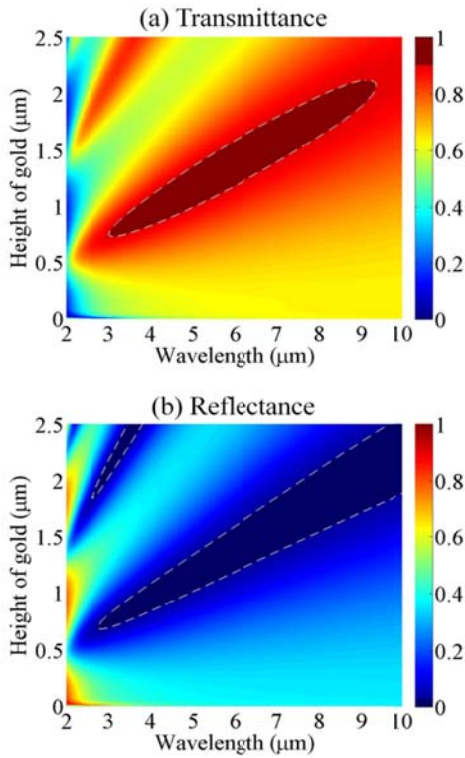

Fig. 6. (a) Transmittance and (b) reflectance versus height of the grating and wavelength. The period is 400 nm and the fill factor is 0.53.

To show the polarization selectivity, transmittance and reflectance from a 1D grating structure with a fill factor of 0.53, a period of 400 nm, and a grating height of 1100 nm, were calculated for both x and y polarization incidence. Transmittance and reflectance for x and y-polarizations are plotted in Fig. 7. For the x-polarization, the 1D grating has a maximal transmission and a minimal reflection at 4.71 μm. For the y-polarization, the 1D structure has zero transmittance and a constant reflectance of 98%, which is due to the effective metallic property of the metamaterial grating in the y-direction.

The results obtained by FDTD simulations were compared with the results calculated from optical wave interference theory using the effective optical constants $\varepsilon_{xx}$ shown in Fig. 2 for normal incidence. The height of the gold grating was chosen to be 1100 nm and the fill factor was chosen to be 0.53. The grating period was chosen to be 50 nm, 100 nm, 200 nm, and 400 nm, respectively. For calculations with the optical wave interference theory, the grating period does not matter because the electric permitivities do not change with the grating period. However, for FDTD simulations, the results vary with the period.

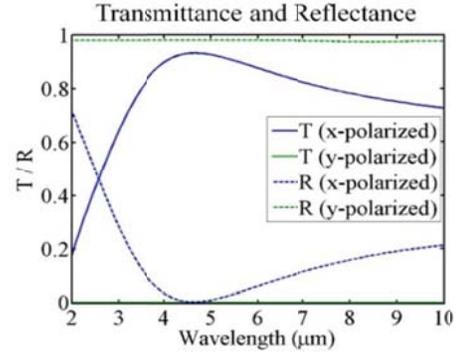

Fig. 7. Transmittance and reflectance versus wavelength for x and y polarizations. The fill factor is 0.53. The period is 400 nm. The height is 1100 nm.

Figures 8(a) and 8(b) show the transmittance and reflectance versus wavelength, calculated with the optical wave interference theory (the solid line) based on the effective medium theory and FDTD simulations (dashed lines). It can be seen that as the grating period decreases, FDTD simulation results are red-shifted and approach the result of the optical interference theory with effective medium material constants. This makes sense because the calculation of the effective electric permittivity assumes that the period is deeply smaller than the wavelength. Thus, the period needs to be very small to closely match the result of optical interference theory. However, it is interesting to find that the optimal anti-reflection structure does not require a deeply small subwavelength period. Finite subwavelength period gratings provide better anti-reflection in 3 − 6 μm range than deeply small subwavelength period gratings do. This property is very significant for fabricating anti-reflection metamaterial coatings, because it is usually technically challenging to fabricate deeply small subwavelength period metal gratings.

To further understand the enhanced transmission through the finite period metal grating structure, we calculated the electric field ($E_x$ and $E_z$ components) and magnetic field distributions ($H_y$) at the peak transmission wavelength of 4.66 μm for a gold grating structure of 400 nm period, 0.53 fill factor, and 1100 nm height. Figs. 9(a) and 9(b) show the electric field profiles of the $E_x$ and $E_z$ components normalized to the electric field amplitude of the incident wave. Fig. 9(c) shows the magnetic field profile of the $H_y$ component normalized to the magnetic field amplitude of incident wave. It can be seen that the incident light is coupled to the propagating surface plasmon polariton wave in the air-gaps of the metal grating. The surface electrical currents on both sides of the metal grating walls oscillate in opposite directions. Propagating surface plasmon wave resonance enhanced transmission through subwavelength narrow metal slits has been discussed earlier in [20-22]. Here, the electric dipoles induced on the dielectric substrate provide a short to the surface plasmon electric current oscillations on the sides of the metal grating walls. Therefore, the metal slits on the dielectric surface support a magnetic resonance at a lower frequency than that of metal slits without a dielectric substrate. The magnetic resonance couples the plasmon-polariton energy to the transmission side of the grating. The coupling between the surface plasmon-polariton wave and the dipoles on the dielectric surface forms the magnetic resonance mode. The electric displacement currents on the side walls of the metal slits and the germanium substrate surface are indicated by the red arrows in Fig. 9(c). The resonance frequency depends on the geometry of the metal grating and also the dielectric constant of the dielectric substrate. Here, the resonance wavelength is about four times the metal grating height. We also calculated light transmission through the same metal grating structure without the germanium substrate. It was found that the magnetic field enhancement is in the middle of the metal grating and the maximal transmission occurs at a shorter wavelength of 2.6 µm.

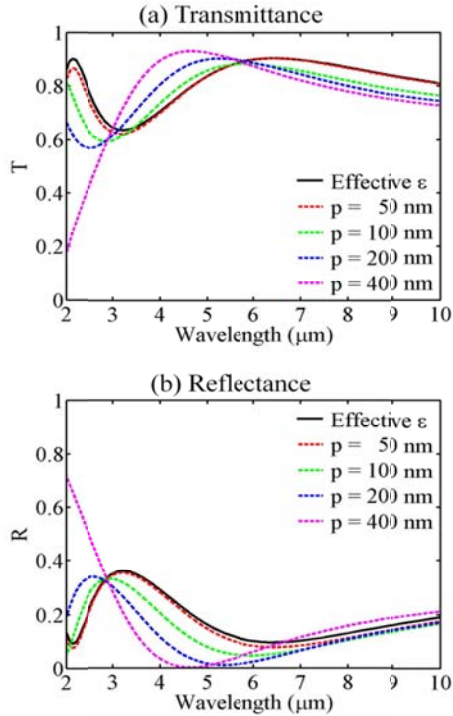

Fig. 8. (a) Transmittance and (b) reflectance versus wavelength for gratings of period: 50 nm, 100 nm, 200 nm, and 400 nm, compared to the transmittance and reflectance calculated by using the effective medium of the structure.

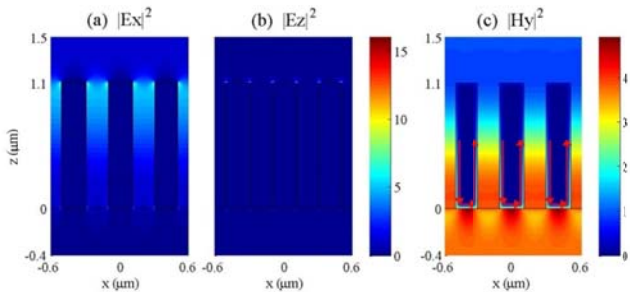

Fig. 9. Normalized electric field profiles of $E_x$ (a) and $E_z$ (b) components at the peak transmission wavelength. (c) Normalized magnetic field profile of $H_y$ component at the peak transmission wavelength.

We also calculated angular dependence of the transmittance and reflectance from an anti-reflective metamaterial grating structure with grating period of 400 nm, fill factor of 0.53, and height of 1.1 µm. Fig. 10 (a) and (b) show the transmittance and reflectance versus wavelength and angle of incidence in the y-z plane. Fig. 11 (a) and (b) show the transmittance and reflectance versus wavelength and angle of incidence in the x-z plane. It was found that good performance anti-reflection occurs within an incident angle range from 0° to 16° (transmission greater than 90%) and from 0° to 12° for reflection smaller than 3%. The performance of antireflection degrades as the angle of incidence increases due to the plasmon coupled optical wave interference inside the thick dielectric-like metamaterial layer.

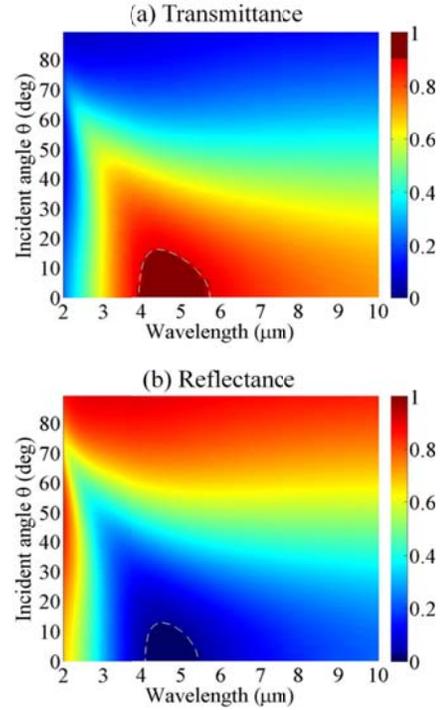

Fig. 10. (a) Transmittance and (b) reflectance versus wavelength and angle of incidence in the y-z plane.

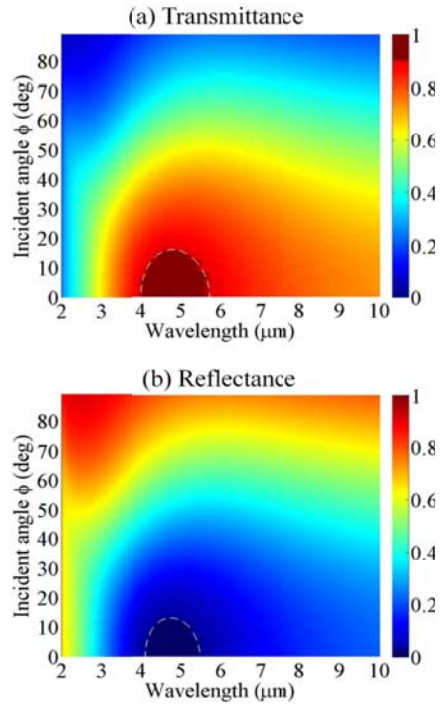

Fig. 11. (a) Transmittance and (b) reflectance versus wavelength for different angles of incidence in the x-z plane.

## 3. SUMMARY

In this paper, a metamaterial structure consisting a 1-D subwavelength metal/air-gap grating was investigated for antireflection coating for germanium substrate. The 1D structure metamaterial functions as an effective dielectric in the polarization direction perpendicular to the metal grating lines with designer electric permittivity. The plasmon-coupled optical wave resonance in the metamaterial structure results in complete elimination of the reflection and enhancement of transmission. It is found that it is not necessary for the grating period to be deeply smaller than the wavelength for achieving antireflection. The thickness of the subwavelength grating metamaterial antireflection coating is approximately a quarter of the free space wavelength. The anti-reflection performance is sensitive to increase of the angle of incidence. Sensitivity to the angle of incidence can be advantageous for imaging systems where unwanted stray light needs to be rejected.

## ACKNOWLEDGEMENTS

This work was partially supported by the National Science Foundation (NSF) through the award No. 1158862. W. Kim acknowledges the support from the Alabama Graduate Research Scholars Program. J. Hendrickson acknowledges the support from the US Air Force Office for Scientific Research (AFOSR) under LRIR No. 12RY05COR and 15RYCOR159.